\documentclass[twocolumn]{webofc}
\usepackage[varg]{txfonts}   
\begin{document}
\title{New Search for Mirror Neutron Regeneration}

\author{\firstname{L. J.} \lastname{Broussard}\inst{1}\fnsep\thanks{\email{broussardlj@ornl.gov}} \and
        \firstname{K. M.} \lastname{Bailey}\inst{1} \and
        \firstname{W. B.} \lastname{Bailey}\inst{1} \and
        \firstname{J. L.} \lastname{Barrow}\inst{2} \and
        \firstname{K.} \lastname{Berry}\inst{1} \and
        \firstname{A.} \lastname{Blose}\inst{3} \and
        \firstname{C.} \lastname{Crawford}\inst{3} \and
        \firstname{L.} \lastname{Debeer-Schmitt}\inst{1} \and
        \firstname{M.} \lastname{Frost}\inst{1,2} \and
        \firstname{A.} \lastname{Galindo-Uribarri}\inst{1} \and
        \firstname{F. X.} \lastname{Gallmeier}\inst{1} \and
        \firstname{C. E.} \lastname{Gilbert}\inst{1,2} \and
        \firstname{L.} \lastname{Heilbronn}\inst{2} \and
        \firstname{E. B.} \lastname{Iverson}\inst{1} \and
        \firstname{A.} \lastname{Johnston}\inst{2} \and
        \firstname{Y.} \lastname{Kamyshkov}\inst{2} \and
        \firstname{P.} \lastname{Lewiz}\inst{2} \and
        \firstname{I.} \lastname{Novikov}\inst{4} \and
        \firstname{S. I.} \lastname{Penttil{\"a}}\inst{1} \and
        \firstname{S.} \lastname{Vavra}\inst{2} \and
        \firstname{A. R.} \lastname{Young}\inst{5}
}

\institute{Oak Ridge National Laboratory, Oak Ridge, TN 37831, USA 
\and
           University of Tennessee, Knoxville, TN 37996  USA 
\and
           University of Kentucky, Lexington, KY 40506  USA
\and
           Western Kentucky University, Bowling Green, KY 42101  USA
\and
           North Carolina State University, Raleigh, NC 27695  USA
          }

\abstract{%
The possibility of relatively fast neutron oscillations into a mirror neutron state is not excluded experimentally when a mirror magnetic field is considered. Direct searches for the disappearance of neutrons into mirror neutrons in a controlled magnetic field have previously been performed using ultracold neutrons, with some anomalous results reported. We describe a technique using cold neutrons to perform a disappearance and regeneration search, which would allow us to unambiguously identify a possible oscillation signal. An experiment using the existing General Purpose-Small Angle Neutron Scattering instrument at the High Flux Isotope Reactor at Oak Ridge National Laboratory will have the sensitivity to fully explore the parameter space of prior ultracold neutron searches and confirm or refute previous claims of observation. This instrument can also conclusively test the validity of recently suggested oscillation-based explanations for the neutron lifetime anomaly.
}
\maketitle
\section{Introduction}
\label{intro}

The astrophysical evidence for dark matter is strong. However, despite decades of searches, we still do not know what the particle nature of dark matter is. With the diminishing available parameter space for popular candidates such as WIMPs (weakly interacting massive particles), the scientific community has recognized the importance of exploring other possibilities~\cite{Battaglieri:2017aum}. In identifying other potential explanations for gravitational observations, a common strategy is to link the dark matter to other questions or anomalies in particle physics. The neutron lifetime puzzle, the nearly 4~$\sigma$ difference between the neutron lifetime measured using cold neutron and ultracold neutron techniques, has motivated searches for mirror and dark neutrons~\cite{Berezhiani:2005hv,Berezhiani:2008bc,Fornal:2018eol}. 

The hypothesis of a neutron transition into a dark state has been significantly constrained by direct experimental limits and neutron star masses~\cite{Tang:2018eln,Sun:2018yaw,Baym:2018ljz,Motta:2018rxp}. Mirror neutron oscillation searches have been performed using ultracold neutrons with some anomalous signals reported~\cite{Serebrov:2008hw,Altarev:2009tg,Berezhiani:2012rq,Berezhiani:2017jkn}. Current limits on oscillations are modest: $\tau > 448$~s assuming no mirror magnetic field~\cite{Serebrov:2008hw}, and $\tau > 17$~s assuming mirror magnetic fields below about 170~mG~\cite{Berezhiani:2017jkn}. An improved search with ultracold neutrons projects a significantly improved sensitivity over this magnetic field range~\cite{Prajwal}, and new ideas with cold neutron beams have been put forth that can resolve the ambiguity associated with the source of ultracold neutron losses~\cite{Berezhiani:2017azg,Broussard:2017yev}.

\begin{figure*}[t]
\centering
\includegraphics[width=\textwidth]{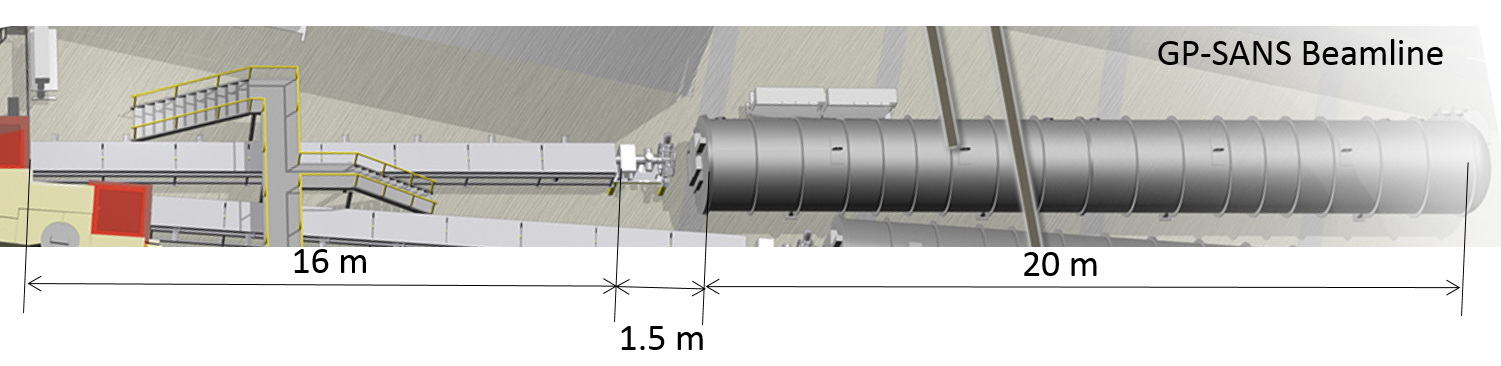}
\caption{The GP-SANS instrument at HFIR includes a 16~m long collimation vacuum tube, 1.5~m long sample region, and 20~m long detector tank.}
\label{fig:gpsans}       
\end{figure*}

\section{Mirror Neutron Oscillations}
\label{osc}

The formalism for the oscillation of neutrons into mirror neutrons has been developed considering the possibility of a mirror magnetic field~\cite{Berezhiani:2005hv,Berezhiani:2008bc,Berezhiani:2012rq}.  The probability for unpolarized free neutrons to oscillate in the presence of a (mirror) magnetic field $B$ ($B'$) is
\begin{align}
P(t) &= \frac{sin^2\left[(\omega-\omega')t\right]}{2\tau^2(\omega-\omega')^2} +  \frac{sin^2\left[(\omega+\omega')t\right]}{2\tau^2(\omega+\omega')^2} \nonumber \\
&+ \left(cos\beta\right) \left( \frac{sin^2\left[(\omega-\omega')t\right]}{2\tau^2(\omega-\omega')^2} -  \frac{sin^2\left[(\omega+\omega')t\right]}{2\tau^2(\omega+\omega')^2}  \right),
\end{align}
where $t$ is the free flight time of the neutron, $\tau$ is the characteristic oscillation time, 
$\omega=\frac{1}{2}|\mu B|$, and $\omega'=\frac{1}{2}|\mu' B'|$ with (mirror) magnetic moment $\mu$ ($\mu'$). 
The probability is enhanced quadratically with longer $t$ and by selecting the magnitude of $B$ approaching the unknown $B'$ and with small angle $\beta$ between them, introducing a resonance in the probability where $\omega \approx \omega'$, assuming $\mu=\mu'$. In a mirror neutron oscillation search using ultracold neutrons, the low neutron density is balanced by the very slow velocities (and therefore long flight times) and many thousands of bounces, and the modest scale of the experiment is convenient for applying a magnetic field with the required level of uniformity and control. Cold neutron beams require long and large area beam guides to achieve useful free flight times, but offer many orders of magnitude higher neutron intensities. With cold neutrons we have a chance to unlock the ability to search for a rarer phenomenon with less ambiguity. By preparing two separated regions with carefully controlled magnetic fields, we can search for neutrons that disappear into mirror neutrons in the first region, pass through a wall, then regenerate into detectable neutrons in the second region. Adding a controlled magnetic field allows a suppression or enhancement of the effect.

\section{Searches at HFIR}
\label{hfir}

The General Purpose Small Angle Neutron Scattering (GP-SANS) instrument at the 85~MW High Flux Isotope Reactor (HFIR)~\cite{CROW2011S71} has been identified as a promising experimental setup for mirror neutron oscillation searches, with intensities of order $10^{10}$~n/s between 4--25~\AA~ expected to be available when operated in white beam mode (Fig.~\ref{fig:gpsans}). The instrument includes a roughly 16~m long vacuum tube used for collimation of the neutron beam. This section is slated for a major upgrade in the second half of 2019, which in particular will enable more rapid change-out of beam guides and removal of magnetic components, and can accommodate larger beam guides of up to 20~cm diameter. These improvements will allow higher sensitivity oscillation studies.

GP-SANS also includes a separate vacuum vessel, the detector tank, which serves as an adjustable-length secondary flight path for neutrons scattered from a sample. Samples are located in a 1.5~m long gap between the collimating section and vacuum window of the detector tank. The ``sample'' for our experiment is actually a high suppression re-entrant beamstop, which absorbs the neutrons and permits passage of only the sterile mirror neutrons. The 20~m long, 2.5~m diameter, cadmium-lined detector tank houses a roughly 1$\times$1~m$^2$ area neutron detector consisting of 192 $^3$He tubes arranged in two staggered planes~\cite{BERRY2012179}. The detector is position sensitive with pixel size of 5.1$\times$4.2 mm$^2$, such that the central $\sim10\%$ in area can be used to search for a regeneration signal, and the outer area can be used for an \textit{in situ} background determination. With the addition of a magnetic field, the collimation vacuum tube and detector tank can serve as a ``disappearance region'' or ``regeneration region,'' respectively, for neutron oscillation searches. We have identified three types of neutron-mirror neutron oscillation searches that can be performed at GP-SANS, which we briefly summarize here.

\subsection{Resonance at low magnetic fields}
\label{lowB}

\begin{figure}[b]
\centering
\includegraphics[width=8cm]{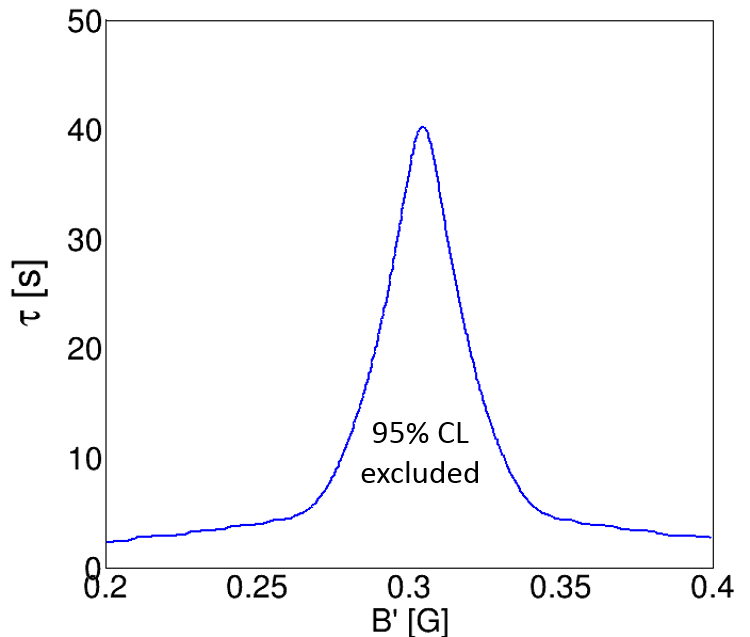}
\caption{Sensitivity of the characteristic oscillation time vs.\ mirror magnetic field after one week regeneration experiment at GP-SANS with an applied magnetic field of 0.3~G.}
\label{fig:exclusion}       
\end{figure}

A sensitivity study of neutron--mirror neutron oscillation searches at GP-SANS has been performed previously~\cite{Broussard:2017yev}. This simulation study included total backgrounds in the region of interest of 0.3~cps, development of a 3D magnetic field control system composed of solenoidal and cosine theta coils with 2~mG uniformity, and scan of magnetic fields simultaneously in both of the separated regions of up to 125~mG in steps of 5~mG, in four directions (forming a tetrahedron) to optimize sensitivity. The results of this study indicates that a search for regeneration of neutrons would exclude $\tau<15$~s over $-125$~mG~$<B'<$~125~mG (90\% C.L.) in a one week experiment (excluding possible systematic studies). GP-SANS may also be used as a powerful cross-check in the event that an anomalous signal is discovered in current ultracold neutron searches, including outside this magnetic field range. An example of the sensitivity of a one week experiment to a search at a single magnetic field magnitude of 0.3~G is shown in Fig.~\ref{fig:exclusion}, where an oscillation time of up to $\tau<40$~s could be probed. This measurement requires large area beam guides instead of collimation in the primary vacuum vessel, and a detailed magnetic field mapping campaign of the detector tank. We note the ultimate sensitivity is expected to be lower than current ultracold neutron efforts~\cite{Prajwal}.

\subsection{Resonance at high magnetic fields}
\label{highB}

Recently, a new idea has surfaced which attempts to explain the neutron lifetime discrepancy using mirror neutron oscillations~\cite{Berezhiani:2018eds}.  This model suggests that the cold neutron beam lifetime experiment at NIST~\cite{Yue:2013qrc} is sensitive to a very high probability of oscillations at much larger magnetic field strengths, of up to 4.6~T, to compensate a small mass splitting of order 100~neV between the neutron and mirror neutron. As neutrons enter the magnet and first cross the compensating field, up to 1\% are found in the mirror state, in which the decay to a mirror proton, which is not trapped by the laboratory magnetic field, is missed. The mirror neutrons which don't decay still have a high probability of being found in the pure neutron state after exiting the magnet, resulting in a too-long measured lifetime.

\begin{figure}[h]
\centering
\includegraphics[width=7cm]{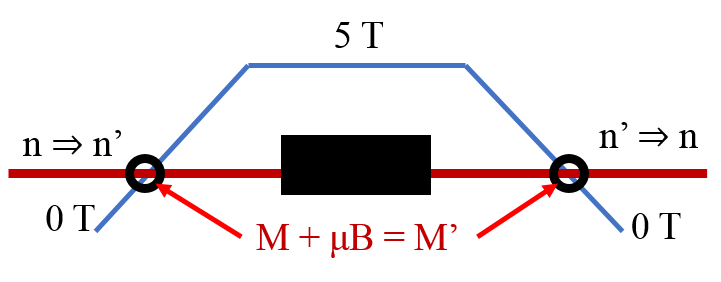}
\caption{Simplified concept for search for mirror neutron disappearance and regeneration at high magnetic field.}
\label{fig:highfield}       
\end{figure}

A simple experiment can be performed to test the hypothesis of neutron oscillation at large (order T) magnetic fields, using a similar disappearance-regeneration style approach (Fig.~\ref{fig:highfield}). A neutron beamstop can be located at the high field region, so that only the neutrons which are found in the mirror state can pass through, and then transform back to the ordinary neutron state after crossing the compensating field, and be detected.

The sensitivity of this type of search at GP-SANS can be estimated using formula 20 of~\cite{Berezhiani:2018eds}. A monochromatic beam of 4.75\AA~(velocity $v=$833~m/s) with $\Delta \lambda/\lambda\sim13\%$ is available with estimated intensity of 10$^7$ n/s, and a 5~T superconducting magnet can be installed in the sample region of the instrument. For a non-adiabatic crossing of the compensating field (at position $z_c$), the probability for a neutron to be found in the mirror neutron state at the beamstop $P_{nn'}\approx \frac{\pi}{4} \xi$ depends on an adiabaticity parameter $\xi$,
\begin{align}
\xi = \Delta m \sin^2 2 \theta_0 v^{-1} \left(\frac{d \ln B(z)}{dz}\biggr\rvert_{z_{c}}\right)^{-1}.
\end{align}
For example, using $\theta_0 \sim 10^{-3}$ as in~\cite{Berezhiani:2018eds},  and supposing the compensating field to be $B=$2~T ($\Delta m=120$~neV), and with $\frac{d \ln B}{dz}\bigr\rvert_{z_c} = 0.2$ from the calculated field map of the superconducting magnet, we find $\xi\sim0.04$ and $P_{nn'}\sim$3\%. Therefore if mirror neutrons are the source of the lifetime discrepancy as suggested in~\cite{Berezhiani:2018eds}, a dramatic effect would be observed, with $10^7\cdot(0.03)^2\sim10^4$ regenerated neutrons per second. 

Much more sensitive limits than needed to explain the lifetime discrepancy can be placed with a reasonably high suppression beamstop. The background from transmission through the beamstop could be subtracted by comparing the observed rate with and without the presence of a large magnetic field. Large changes in neutron backgrounds from scattering from other instruments in the facility can be monitored using detectors outside the beamline, and sensitivity to background can be reduced by positioning the GP-SANS detector further back in the shielded detector tank. Because the oscillation probability depends strongly on the neutron velocity, a useful cross-check of any unexplained magnetic field-dependent effect can quickly be obtained by selecting different neutron wavelengths using the GP-SANS velocity selector. This kind of experiment can be performed quickly, involving only the development of a beamstop with $<$10$^{-6}$ transmission, required for all future stages of oscillation searches. 

\subsection{Transition magnetic moment}
\label{tran}

An alternative model for mirror neutron oscillations has recently been suggested where the neutron and mirror neutron have an additional magnetic moment, a Transition Magnetic Moment (TMM) $\eta$ of common magnitude, which enters in the non-diagonal part of the Hamiltonian operator that couples the neutron and mirror neutron correspondingly with the ordinary and mirror electromagnetic fields~\cite{ntrans}. In this model, if both $B$ and $B'$ are above a few mG, the n$\rightarrow$n' transition probability is constant and doesn't depend on the magnitude of the magnetic field, but only on the magnitude of $\eta$. This model can shed light on the neutron lifetime anomaly, in which the ultracold neutron bottle lifetime experiments are susceptible to an unaccounted loss due to oscillations into mirror neutrons in the volume. In a material or magnetic bottle, the neutron has a chance to collapse into the mirror neutron state (and be lost) at each bounce from the material walls or when accelerated by magnetic gradients. A n$\rightarrow$n' transition probability of 10$^{-7}$--10$^{-9}$ due to the TMM of order $\eta/\mu \sim$10$^{-4}$--10$^{-5}$ could explain the lifetime discrepancy.

A sensitive search for a TMM can be performed by using an applied magnetic field gradient to more rapidly force a decoherence of the entangled ordinary and mirror neutron states. The length scale of the decoherence $\Delta x$ for neutrons passing through a magnetic field gradient can be estimated using the relation (adapted from formula 17 of~\cite{ntrans})
\begin{align}
\frac{\Delta B}{\Delta x} &\sim \frac{\hbar v }{\mu (\Delta x)^2}
\end{align}
By maximizing the gradient, the number of decoherence lengths $\Delta x$ that the neutron passes through is maximized.  Each $\Delta x$ represents an opportunity for a n$\rightarrow$n' transition.

\begin{figure}[h]
\centering
\includegraphics[width=7cm]{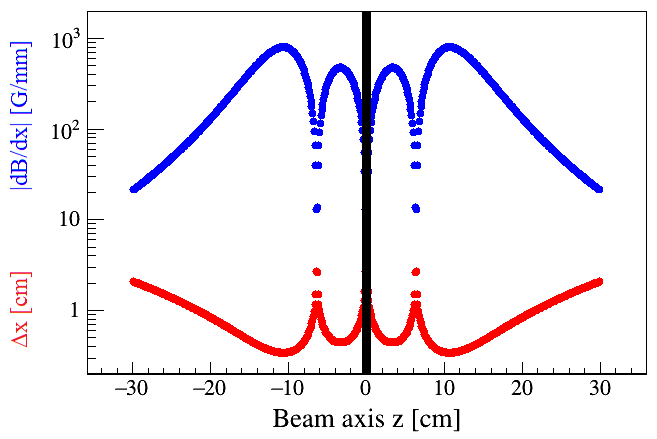}
\caption{Magnetic field gradient (blue) and decoherence length (red) for 4.75~\AA~neutrons in a superconducting magnet available at GP-SANS used to calculate sensitivity for a neutron TMM search. The location of the beamstop is indicated by the black vertical band.}
\label{fig:TMM}       
\end{figure}

A preliminary search for a neutron TMM can be performed in a simple regeneration experiment using a beamstop inside the superconducting magnet at GP-SANS. The calculated magnetic field gradient and corresponding decoherence length along the neutron beam axis inside the magnet is plotted in Figure~\ref{fig:TMM}. The unusual magnetic profile is due to the split coil which accommodates a perpendicular bore to allow insertion of samples in the magnet center. The 4.75~\AA~neutron beam will pass through roughly 450 decoherence lengths on each side of a beamstop centered in the magnet, enabling a sensitivity of up to $\eta/\mu < 2\times10^{-4}$ to be probed with a beamtime allocation of $\sim1$ day.  With the velocity selector removed and operating in white beam mode, this limit can be improved to $\eta/\mu < 8\times10^{-5}$.

Large gradients over longer flight paths could also be implemented using a series of coils with alternating currents as suggested in~\cite{ntrans} or with a cylindrical Halbach array composed of permanent magnets to create strong longitudinal or radial gradients. Gradients of about 2 kG/m could be constructed in this way, enabling a sensitivity to a n$\rightarrow$n' transition probability approaching 10$^{-9}$ or a TMM of $\eta/\mu \sim10^{-5}$.  An alternate scheme that balances the magnetic potential and Fermi potential of a gas has also been described which enable even higher sensitivity~\cite{ntrans}.

To achieve the sensitivity required for a TMM search, a cold neutron beamstop with a transmission rate of less than 1 in 10$^{12}$ is required.  Using the PHITS~\cite{phits} transport code, we estimate that 1~cm of boron carbide B$_4$C (about 20$\%$ natural abundance of $^{10}$B) is sufficient for the neutron spectrum available at GP-SANS. The beamstop will be implemented using a re-entrant design to limit backscattering. To characterize fluctuating neutron backgrounds in the facility, due to changing conditions of instruments on other beamlines, a secondary detector of 8 cylindrical $^3$He gas chambers with similar geometry to the GP-SANS primary detector has been installed outside the detector tank.

\section{Conclusions}

Mirror neutron oscillations have been suggested as a solution to the neutron lifetime puzzle and if discovered would provide a clue as to the particle nature of dark matter. A diverse program of searches for neutron oscillations into sterile mirror neutrons is proposed using the GP-SANS instrument at HFIR.  These cold neutron regeneration searches would be complementary to ultracold neutron storage measurements and avoid the ambiguity associated with sources of ultracold neutron losses. With minimal impact to the instrument and with reasonably short running times, we can definitively confirm or refute that an oscillation with a small mass splitting exists at a level sufficient to explain the lifetime discrepancy, and search for the existence of a TMM with sensitivity of $\eta/\mu \sim 10^{-5}$. With additional instrumentation, we can search for an oscillation time of up to $\tau<40$~s due to a resonance at low magnetic fields. A planned upgrade to GP-SANS will enable further improvements in sensitivity of the mirror neutron oscillation searches.

\section{Acknowledgements}

This research was sponsored by the Laboratory Directed Research and Development Program [project 8215] of Oak Ridge National Laboratory, managed by UT-Battelle, LLC, for the U. S. Department of Energy, and was supported in part by the U.S. Department of Energy, Office of Science, Office of High Energy Physics [contract {DE-SC0014558}] and Office of Nuclear Physics [contracts {DE-AC05-00OR2272}, {DE-SC0014622}, {DE-SC0008107}, {DE-FG02-ER41042}, and {DEFG02-03ER41258}], and in part by the National Science Foundation [contract PHY1615153].


\bibliography{main.bib}

\end{document}